\documentclass{cup-hpl}

\usepackage{upgreek}
\usepackage{siunitx}
\usepackage{libertine} 
\usepackage{graphicx}
\usepackage{dcolumn}
\usepackage{bm}
\usepackage{float}
\usepackage{natbib}

\begin{document}

\shorttitle{High repetition rate targets and optics from liquid microjets}                                   
\shortauthor{K. M. George et al.}
\title{High repetition rate ($\geq$ kHz) targets and optics from liquid microjets for the study and application of high intensity laser-plasma interactions}

\author[1]{K. M. George \corresp{kevin.george.8.ctr@us.af.mil}}
\author[1]{J. T. Morrison}%
\author[2,3]{S. Feister}%
\author[1]{G. Ngirmang}%
\author[4]{J. R. Smith}%
\author[4]{A. J. Klim}%
\author[5]{J. Snyder}%
\author[4]{D. Austin}%
\author[1]{W. Erbsen}%
\author[1]{K. D. Frische}%
\author[7]{J. Nees}%
\author[4]{C. Orban}%
\author[4,8]{E. A. Chowdhury}%
\author[9]{W. M. Roquemore}%

\address[1]{Innovative Scientific Solutions, Inc., Dayton, Ohio 45459, USA}%
\address[2]{Department of Physics and Astronomy, University of California Los Angeles, Los Angeles, CA 90095, USA}%
\address[3]{Computer Science Program and Applied Physics Program, California State University Channel Islands, Camarillo, California 93012, USA}%
\address[4]{Department of Physics, Ohio State University, Columbus, Ohio 43210, USA}%
\address[5]{Department of Mathematical and Physical Sciences, Miami University, Hamilton, Ohio 45011}%
\address[6]{Center for Ultrafast Optical Sciences, University of Michigan, Ann Arbor, Michigan 48109, USA}%
\address[7]{Intense Energy Solutions, LLC., Plain City, Ohio 43064, USA}%
\address[8]{Air Force Research Laboratory, WPAFB, Ohio 45433, USA}%

\begin{abstract}
High intensity laser-plasma interactions produce a wide array of energetic particles and beams with promising applications. Unfortunately, high repetition rate and high average power requirements for many applications are not satisfied by the lasers, optics, targets, and diagnostics currently employed. Here, we address the need for high repetition rate targets and optics through the use of liquids. A novel nozzle assembly is used to generate high-velocity, laminar-flowing liquid microjets which are compatible with a low-vacuum environment, generate little to no debris, and exhibit precise positional and dimensional tolerances.  Jets, droplets, submicron thick sheets, and other exotic configurations are characterized with pump-probe shadowgraphy to evaluate their use as targets. To demonstrate a high repetition rate, consumable, liquid optical element, we present a plasma mirror created by a submicron thick liquid sheet. This plasma mirror provides etalon-like anti-reflection properties in the low-field of $0.1\%$ and high reflectivity as a plasma, $69\%$, at a repetition rate of \SI{1}{\kilo\hertz}. Practical considerations of fluid compatibility, in-vacuum operation, and estimates of maximum repetition rate in excess of 10 kHz are addressed. The targets and optics presented here enable the use of relativistically intense lasers at high \textit{average} power and make possible many long proposed applications. 

\end{abstract}


\maketitle

\section{Introduction} \label{introduction}

Relativistically intense laser-plasma interactions are capable of generating energetic sources of radiation and particles. X-ray, gamma ray, electron, positron, proton, heavy ion, and neutron sources stemming from these interactions have all been created and characterized \cite{TNSA_Hatchett,TNSA_Snavely,Esirkepov2004,Kar2008,Yin2007,Kaw1970,Sprangle1987,Guerin1996,Petrov2009,Bartal2012,Ceccotti2007,Schwoerer2006,Gordon1994,jiang2016microengineering,jiang2014enhancing,Link2018,Mackinnon2001,Hegelich2002}. These sources are advantageous for a range of applications, due to the small source sizes \cite{Borghesi2004,Schreiber2004,Glinec2005} and short time durations \cite{Murnane1991} which are unmatched by conventional techniques. Additionally, a single, table-top laser system can be used to generate a wide array of energetic particles and beams \cite{Mourou1997,Hooker2013}.

Facilities capable of reaching relativistic intensities have been available at laboratories around the world for decades. Using these systems, numerous electron and ion acceleration mechanisms were discovered including: target normal sheath acceleration \cite{TNSA_Hatchett,TNSA_Snavely}, light-sail \cite{Esirkepov2004,Kar2008}, breakout afterburner \cite{Yin2007}, and radiation pressure acceleration \cite{Kaw1970,Sprangle1987,Guerin1996}. Complementary to these studies is the optimization of target element \cite{Petrov2009}, shape \cite{Bartal2012}, thickness \cite{Ceccotti2007}, use of microstructured surfaces \cite{Schwoerer2006,Gordon1994,jiang2016microengineering,jiang2014enhancing}, and tailoring the preplasma scale length \cite{Link2018,Mackinnon2001,Hegelich2002} which serve to enhance the particle source total number, divergence, and peak energy to meet the needs of various applications. 

Of particular interest to the application of these sources is use in proton cancer therapy \cite{Bulanov2002,Bulanov2002b}, neutron generation \cite{Disdier1999,Norreys1998}, and nuclear activation \cite{Cowan1999}. Unfortunately, current generation laser, target, optic, and diagnostic techniques do not meet the high repetition rate or high average power needs of these applications. Most relativistically intense laser systems today operate at low repetition rates ranging from $1$ shot per hour to $1$ shot per minute and provide average powers of less than \SI{1}{\watt} \cite{Danson2015}. 

To address these deficiencies and satisfy the requirements of numerous applications, advances in laser technology are being implemented to construct new facilities which operate with high average power and at high repetition rates \cite{Haefner2017,Sistrunk2017,Rus2017}. These lasers run at between $1$ and \SI{10}{\hertz} and provide high average power in excess of \SI{100}{\watt}. Development of future road maps for petawatt-class lasers operating at kHz and higher repetition rates for advanced accelerator concepts and high energy particle sources are already underway \cite{Leemans2017}. To use these current or future high repetition rate systems to their full capability: target, optic, and diagnostic technology, along with new operational techniques must be developed \cite{Prencipe2017}.

Target systems designed for low repetition rate operation typically rely on solid metal foils which are individually rastered or rotated into place and aligned to within few micron accuracy before irradiation. At a $1$ to \SI{10}{\hertz} repetition rate this process is feasible, but scaling to even higher repetition rates quickly becomes untenable. The total number of targets per carrier is typically limited to a few hundred or thousand at most, requiring downtime to reload. 

Before alignment, each target must be precisely fabricated and characterized. Use at high repetition rate requires tens to hundreds of thousands of targets for sustained operation throughout the course of just one day. Current fabrication and metrology approaches are not suited to meet the over thousand-fold increase in demand for these targets.

Another important topic to consider, as repetition rates increase, is technology to improve the properties of the ultra-intense laser pulse before it arrives at the target. Plasma mirrors, a commonly employed optical element, aim to improve the temporal pulse contrast of the laser and prevent the deleterious generation of preplasma \cite{Kapteyn1991}. Typically comprised of a dielectric anti-reflection coating on an optical quality substrate, plasma mirrors are one-time-use optics in which the irradiated region is destroyed on each laser shot. Large area plasma mirrors are commonly used and rastered for multiple exposures in order to limit cost of such devices, however, use in $1$ to \SI{10}{\hertz} systems is impractical purely from a cost standpoint.

High repetition rate operation presents new operational challenges not present with low repetition rate systems. As laser intensities and peak powers increase, combined with high repetition rate operation, the potential for debris accumulation and damage to sensitive optics increases. Exceedingly expensive final focusing optics may need to become consumable, or protected by consumable pellicles, when operating in these environments. Lower cost, lower quality disposable focusing optics or plasma optics have been proposed as substitutes \cite{Kirkwood2018}.

For all of these reasons there is now a consensus that much more work needs to be done to address these concerns (Prencipe \textit{et al.} 2017). Earlier pioneering works had the foresight to identify and undertake many of these issues \cite{thoss2003kilohertz,thossthesis}. In doing so, an intense, kHz repetition rate, femtosecond laser was integrated with a liquid jet target for developing integrated sources of radiation and particles. Given recent emphasis on, and developments of, relativistically intense, high repetition rate lasers, we bring new insights and results. This work details a high repetition rate mode of operation for targets and optics based on liquid microjets for the application and study laser-plasma interactions. 

Here we present a novel target generation scheme based on high-velocity, laminar-flowing, liquid microjets which support estimated repetition rates up to \SI{40}{\kilo\hertz}. The targets include a \SI{33}{\micro\meter} diameter cylindrical jet, \SI{21} and \SI{55}{\micro\meter} diameter droplets, submicron thick sheets, and other exotic configurations all from a simple and robust nozzle assembly. High repetition rate, consumable, optical elements are demonstrated with a plasma mirror generated by use of a submicron thick liquid sheet. Operating at a \SI{1}{\kilo\hertz} repetition rate in the low-field, the etalon-like anti-reflection properties provides a reflectivity of $0.1\%$. When incident by an intense laser pulse, the triggered plasma reflectivity is $69\%$. We detail our efforts to practically achieve continuous operation in a low vacuum environment, addressing fluid compatibility and maximum proposed repetition rates for each target type.

The paper is organized as follows. We begin by discussing the physics involved in the formation of liquid jets including laminar flow conditions, limitations to the free laminar flow propagation, subsequent breakup effects, and provide characteristic lengths and timescales for the jets formed in this work. The assembly used to generate these liquid microjets is then detailed in Section \ref{liquidmicrojetassembly}, covering the simple and robust nozzle design along with the fluid pump employed. Section \ref{liquidtargets} examines liquid targets: cylindrical jets, droplets, submicron thick sheets, and other geometries. Here, dimensional and positional stability, critical to experimental use, are characterized with short pulse probe beam shadowgraphy. Section \ref{liquidoptics} covers the experimental demonstration of a liquid plasma mirror. We then discuss a number of practical considerations in the design, use, and implementation of this system including: vacuum operation, fluid compatibility, and estimates for the maximum repetition rate. Lastly, the paper is concluded with a discussion of the implications and impact of this work in relation to advancing the capabilities of high repetition rate relativistically intense laser-plasma interactions.

\section{Physics of Liquid Jets} \label{physicsofliquidmicrojets}

The liquid targets and optics described in this work are based on the physics of liquid jets which were first studied in detail by Lord Rayleigh over 100 years ago \cite{Rayleigh1878}. Since that time, our collective understanding of the fundamental physical interactions of liquid jets has enabled widespread use in a range of applications from jet engine propulsion \cite{Ryan1995} to x-ray spectroscopy \cite{Wilson2001}. Here, we briefly address the physics which forms the basis of the target and optic work described later in this paper.

The fundamental component of the target system is a high-velocity, laminar-flowing, liquid microjet. Formation of a continuous, laminar-flowing liquid jet requires that the Reynolds number, $R_\mathrm{e}$, as defined by $R_\mathrm{e} = \rho v d / \eta$ where $\rho$ is the density, $v$ the velocity, $d$ the diameter, and $\eta$ the viscosity, be less than \num[group-separator = {,}]{2000} \cite{Rayleigh1878,Eggers2008}. Flow conditions exceeding this laminar limit generates turbulent instability in the jet leading to premature breakup during propagation \cite{Darbyshire1995,Reynolds1883,Faisst2004}.

Even within this laminar limit, jets are inherently unstable due to the Plateau-Rayleigh instability \cite{Rayleigh1878}. Perturbations in the flow ultimately lead to a minimization of the surface energy which drives breakup of the jet into droplets. This effect occurs over a characteristic distance, the spontaneous breakup length, $L = 12v ( \sqrt{\rho d^3 / \sigma}+ 3 \eta d / \sigma)$ where $\sigma$ is the surface tension \cite{Van2010,Kalaaji2003}. When uncontrolled, the resulting jet decomposes into a droplet spray with largely varying droplet volume and velocity distribution. This effect is detrimental to the practical application of fluid jets, as a long, stable, propagation distance is required in order to permit optical and diagnostic access to the laser-target interaction region.

In some cases, droplet formation in a repeatable manner is desired. One can provide a droplet for every laser pulse by seeding the Pleateau-Rayeigh instability through vibrations or pressure fluctuations to initate droplet formation with high repeatability. When operated in this mode, droplets are formed at repeatable intervals with well-controlled volume and velocity distributions. Additionally, the flow-dependent droplet formation frequency allows for the creation of a droplet train at repetition rates greater than \SI{100}{\kilo\hertz}.

Following from the Plateau-Rayleigh instability, the growth rate of a perturbation to a flowing liquid jet is maximal at the point where $kR_0 \approx 0.697$ with $k$ being the perturbation wavenumber defined by $k = 2 \pi / \lambda$ and $R_0$ equal to the radius of the unperturbed liquid jet.  Given a flow rate of the liquid jet, $r_f$, the spontaneous droplet frequency, $f_\mathrm{d}$ is given by $f_\mathrm{d} = 0.35 r_\mathrm{f} / (\pi^2 R_0^3)$. Utilizing the droplet frequency, the droplet size, $D_\mathrm{d}$, is estimated to be $D_\mathrm{d} = 3.78 R_0$. Droplets smaller than $D_\mathrm{d}$, called satellite droplets, may also be formed during droplet breakoff and appear alternating with the primary droplets in the droplet train \cite{Pimbley1977}.

To give a scale of these values for the parameters used in this work we operate with a nominally \SI{30}{\micro\meter} diameter jet with a controlled fluid velocity of \SI{24}{\meter\per\second}. For ethylene glycol and its associated surface tension and viscosity values, $R_\mathrm{e} = 44.7$ which is well within the laminar flow regime. The spontaneous breakup length, $L$, is \SI{15.8}{\milli\meter}. When operated in mode of droplet formation the natural droplet size is \SI{56.7}{\micro\meter} with a spontaneous droplet frequency of \SI{178}{\kilo\hertz}. 

\section{Liquid Microjet Assembly} \label{liquidmicrojetassembly}
Fundamental to the formation of liquid microjets in this work is our effort to adhere to ease of setup and maintenance. Therefore, many of the components in the microjet assembly are commercial off-the-shelf items which are supplied in large quantity for relatively low cost. The following section details the components of the microjet assembly and construction techniques.

\begin{figure}[t]
\includegraphics[width=\linewidth]{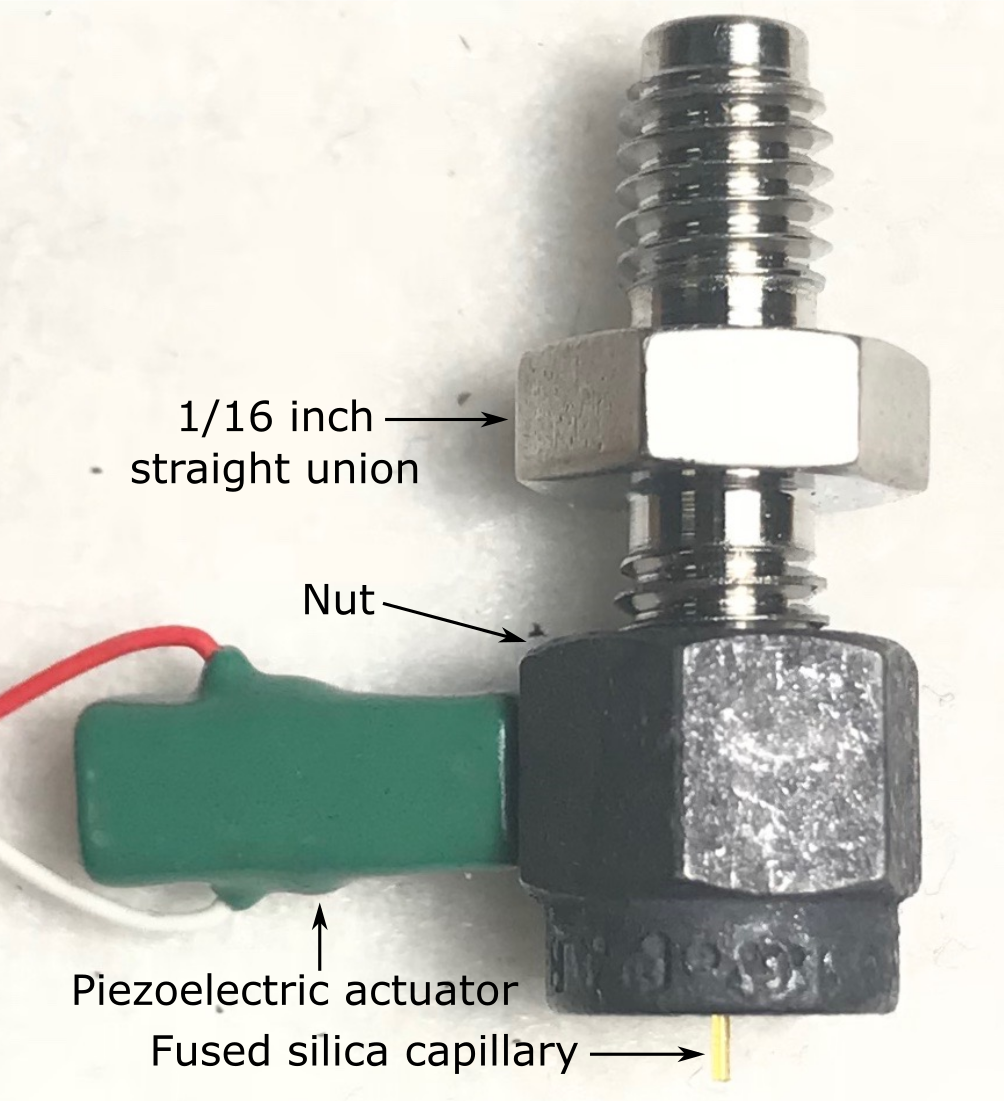}
\caption{Liquid microjet nozzle assembly composed of a 1/16 inch Swagelok fitting, Vespel ferrule, \SI{30}{\micro\meter} inner diameter glass capillary tube, and locking nut with affixed piezoelectric actuator for droplet formation.}
\label{fig:nozzle_and_jet}
\end{figure}

The fundamental components of the liquid microjet nozzle assembly is shown in Figure \ref{fig:nozzle_and_jet}. The nozzle assembly depicted here is comprised of a Swagelok $1/16$ inch straight union, nut, Vespel ferrule, and fused silica capillary. A piezoelectric actuator is affixed with epoxy to the Swaglok nut for the purpose of enabling seeded droplet formation. 

The Swagelok components are standard off-the-shelf items which require no modification for installation in this system. Vespel ferrules are specified to accomodate the outer diameter of the glass capillary, used as the nozzle tip, at \SI{400}{\micro\meter}. The glass capillary is available for purchase by the meter with an outer diameter of \SI{360}{\micro\meter} and available inner diameters ranging from \SI{5} to over \SI{200}{\micro\meter}. A Shortix fused silica capillary cutter is used to cut the capillary, which is then installed into the ferrule and Swagelok assembly by hand. Damaged or clogged nozzles can be cheaply and easily swapped out with the only consumable items being the capillary and Vespel ferrule.

For this work we employ a glass capillary with \SI{30}{\micro\meter} inner diameter. We found nozzles smaller than \SI{25}{\micro\meter} require additional complications of meticulous component inspection, cleaning, and use of multiple stages of sintered steel filters along the fluid lines in order to prevent clogging. Smaller nozzle apertures also require pumps rated for pressures higher than \num[group-separator = {,}]{10000} PSI. Operating above this pressure threshold requires specialized valves, filters, fittings, and lines throughout the system. These complications and clogging failure mode work against the desired simple construction and robust operation of the presented system.

The high-velocity, laminar-flowing, liquid microjet from the nozzle assembly is driven by a high-pressure syringe pump, Teledyne ISCO Model 100 DX, which provides a continuous flow at up to \num[group-separator = {,}]{10000} PSI. The stable pressure control and vibration-free output of the syringe pump makes it particularly well-suited for the formation of laminar, liquid microjets. Changes in the fluid pressure supplied to the nozzle leads to variations in the laminar flow conditions essential for consistent jet formation. Further, pressure waves within the fluid lines can directly couple to vibrations at the nozzle tip which reduces alignment precision and can seed instabilities in the fluid flow causing disintegration of the jet before the expected breakup distance. 

The liquid supply for the syringe pump is maintained at atmospheric pressure and fed into the \SI{100}{\milli\liter} syringe pump reservoir through a \SI{2}{\micro\meter} sintered steel filter to remove particulates. From this reservoir the fluid is transported to the target chamber, at high pressure, via standard $1/16$ inch Swagelok stainless steel tubing and fittings. The fluid line is connected to the nozzle assembly from Figure \ref{fig:nozzle_and_jet}. Under the appropriate flow conditions, a high-velocity, laminar, liquid microjet is formed at the termination of the nozzle assembly capillary and used to form the targets and optics detailed in the following sections.

\section{Liquid Targets} \label{liquidtargets}

The demands for target requirements for laser-plasma interaction (LPI) studies and applications have been discussed and detailed throughout the literature \cite{Prencipe2017}. To review, targets must be thin, of order \SI{10}{\micro\meter} or less, and allow for variable thickness capability, as thin as \SI{10}s of nm, for optimization of certain physical mechanisms such as breakout afterburner or radiation pressure acceleration. Due to fast-focusing optics, alignment along the optical axis must be maintained within few-micron tolerances. Lastly, the target must operate at a low ambient pressure to prevent nonlinear phase effects from impacting the beam propagation to the target \cite{Monot1992,Sullivan1994}. Additional constraints imposed for high repetition rate targets are debris-free operation and low cost per shot. 

Previous efforts to operate with solid density targets at high repetition rates include ribbon spools \cite{Nayuki2003,Noaman2017} and rotating disks \cite{Mordovanakis2010,Borot2012}. These targets are typically thicker than \SI{10}{\micro\meter} and thus are not capable of optimizing the most well-studied ion acceleration process, rear surface target normal sheath acceleration. Both ribbons and disks are not well-suited for continuous, long-term operation due to limited surface area and debris generation. Additionally, ribbon targets lack the positional stability needed for use with fast-focusing optics.

Liquid targets display substantial benefits to high repetition rate operation; the target material can be recycled, they generate little to no debris, but typically require non-negligible operating pressures (e.g. Morrison \textit{et al.} 2015). Application of the appropriate dual pump scheme permits uninterrupted, continuous operation for hours or days.

Liquid sprays or mists have been employed in LPI studies \cite{Mountford1998}, but do not reach the densities required to reflect optical or near-infrared light. Recently, cryogenic hydrogen microjets have demonstrated multi-MeV TNSA of protons at \SI{1}{\hertz} repetition rate \cite{Kim2016,Gauthier2017}, but lacks precise positional control and requires long cool-down times due to cryogenic operation. Liquid crystal films have also exhibited substantial benefits, notably: high vacuum compatibility due to the low fluid vapor pressure, planar geometry, and controllable thicknesses. To date, development of the liquid crystal targets has been performed for few Hz operation, but further work is required to improve film-to-film thickness repeatability and \SI{10}{\hertz} or higher repetition rate capability \cite{Poole2014,Poole2016moderate}.

The remainder of this section will present our results for creating various liquid targets using either water or ethylene glycol and operating at \SI{1}{\kilo\hertz}. For the purpose of characterization, the presented targets were imaged by probe beam illumination. The laser source used is a Coherent Legend which is frequency doubled to \SI{420}{\nano\meter} and has a full width half maximum (FWHM) pulse duration of \SI{80}{\femto\second}. The imaging objective is a Mitutoyo Plan APO Infinity Corrected Long Working Distance 10x microscope objective which is projected with an eyepiece lens onto an ImagingSource DMK 42BUC03 CCD. The spatial resolution of the imaging system was approximately \SI{1}{\micro\meter}. Further details on the laser source and imaging systems can be found in Feister \textit{et al.} 2014 \cite{Feister2014}. 

A set of 1,000 images was recorded for each target type. Image analysis was conducted to determine the size, major and minor axis dimensions, and probability of target presence which accounts for both positional and dimensional fluctuations from target to target. 

Water, at atmospheric pressure, was used for the target generation and characterization in all cases but that of the liquid sheet target. For this case, we employ ethylene glycol, in a vacuum environment. The liquid microjet system is compatible with a range of fluids, which all function in a similar manner, according to the individual fluid properties. To address the difference between the operation of liquid microjets at air versus in vacuum, other current works have found that the resulting microjet properties do not appreciably differ between the two cases \cite{Ekimova2015,Galinis2017}.

\subsection{Liquid Jet Targets} \label{liquidjettarget}
\begin{figure}
	\includegraphics[width=\linewidth]{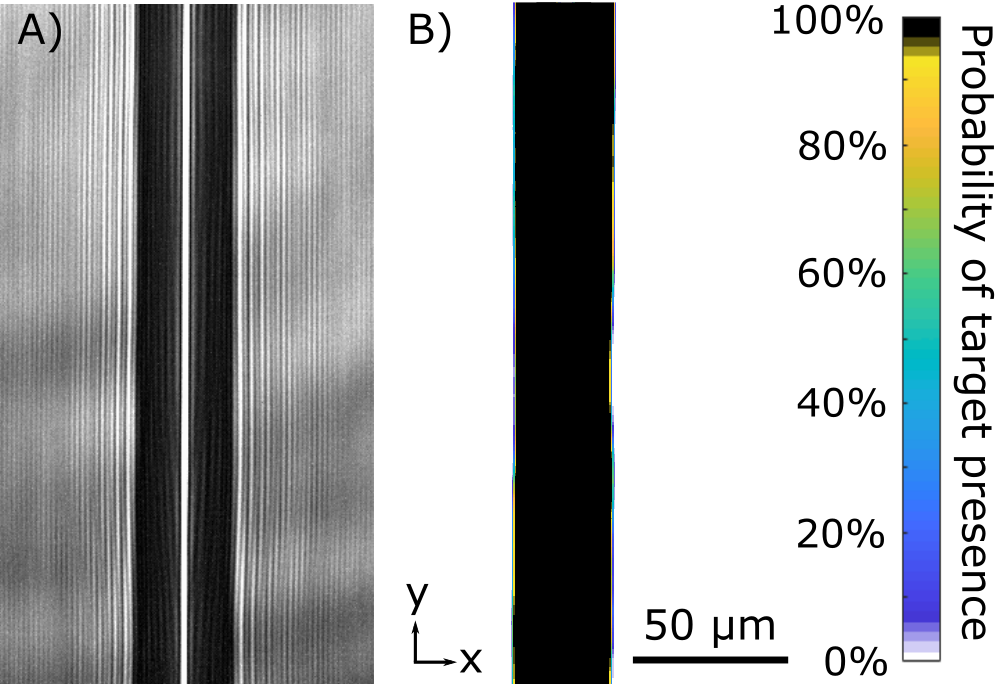}
	\caption{A) Shadowgraphic microscope image of the liquid jet target. A \SI{30}{\micro\meter} inner diameter capillary generates a \SI{33}{\micro\meter} diameter jet. B) False color image of target presence probability. The color scale displays the probability the target appears in the given location over the 1000 target exposures. Black indicates $100\%$ probability the target appears in the given location while white illustrates a $0\%$ probability. The sharp gradient between black and white, shown here, indicates the high positional stability of the liquid jet target.}
	\label{fig:columntarget}
\end{figure}

The most fundamental and simplistic target to generate with the presented system is that of a liquid jet. Figure \ref{fig:columntarget}A) displays an image of the jet target. Here, a capillary with \SI{30}{\micro\meter} inner diameter was used with water to form a cylindrical jet. 

The flow rate from the syringe pump was set to \SI{1}{\milli\liter\per\minute} which generates a fluid velocity of \SI{23.6}{\meter\per\second}. The corresponding Reynolds number of $795$ places the jet well within the limit for laminar flow. The spontaneous breakup length for this condition is \SI{5.76}{\milli\meter} which allows for laser and diagnostic field of view access to the target interaction region.

As previously mentioned, characterization of the jet was performed with short pulse, microscope shadowgraphy images. Image analysis was conducted to quantify the size and stability of the jet. We find that the diameter of the liquid jet is \SI{33}{\micro\meter} with a $1\sigma$ standard deviation in the diameter of the jet from shot-to-shot at less than the \SI{1}{\micro\meter} resolution of the imaging system.

The positional stability of the jet was assessed by means of the same image analysis routine. 
From the \num[group-separator = {,}]{1000} images collected, we identify the region of the image where the target is located. We then calculate on a per-pixel basis, the probability that the target will appear within that pixel over the \num[group-separator = {,}]{1000} recorded instances. We refer to this metric as the probability of target presence, best illustrated by Figure \ref{fig:droplettargets}.

While this metric does not necessarily quantitatively describe the size, shape, and position of the targets due to the convolution between these three variables, it does provide an instructive and qualitative indication of the target stability. Note that black indicates that for all \num[group-separator = {,}]{1000} occurrences a portion of the target was located in that position while white shows where the target does not appear. 

The resulting probability of target presence image for the jet target is shown in Figure \ref{fig:columntarget}B). The standard deviation in the position of the left and right edges of the jet are again better than the \SI{1}{\micro\meter} resolution of the imaging system. This stability is attributed to the mechanical stability of the nozzle holder and consistently pressure and flow rate provided by the syringe pump.

With regards to the applicability of the jet target, while not ideal for electron and ion acceleration due to the circular cross section, this particular target has found use due to the simplicity and straightforward implementation. Initial studies of intense laser-liquid interactions, by Thoss \textit{et al.} 2003 sought to develop sources from a Ga liquid jet with \SI{30}{\micro\meter} diameter irradiated by a \SI{1}{\kilo\hertz} repetition rate, \SI{50}{\femto\second} pulse duration, laser at an intensity of \SI{3e16}{\watt\per\centi\meter\squared} \cite{thoss2003kilohertz}. 

More recent work using the nozzle assembly described in this work has been performed. Backward moving electron acceleration far exceeding ponderomotive scalings at \SI{1}{\kilo\hertz} repetition rate with relativistic intensities were demonstrated \cite{orban2015backward,backward2016,Morrison2015,Feister2017}. These works were performed with water at tens of Torr background pressure, though other works have conducted experiments with the use of water jets in vacuum at far lower pressures \cite{Stan2016}. Overall, the jet target provides a simple and robust starting point for the presentation of liquid targets for high intensity LPI experiments and applications.

\subsection{Liquid Droplet Targets} \label{liquiddroplettargets}

\begin{figure*}[t]
	\includegraphics[width=\linewidth]{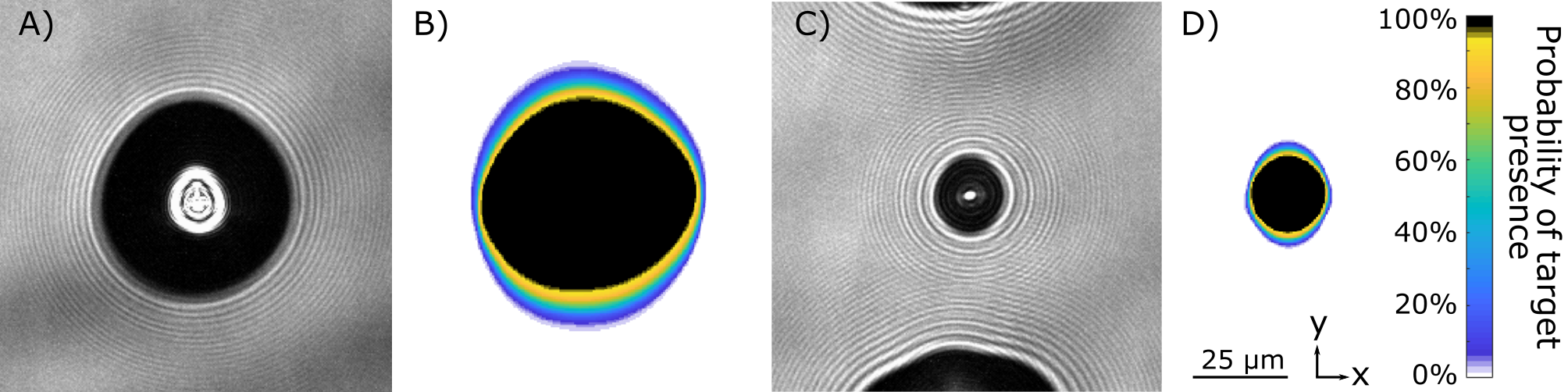}
	\caption{A) \& C) Shadowgraphic microscope images of primary and satellite droplet targets formed by manipulation of the piezoelectric actuator attached to the liquid jet nozzle. The primary droplet in A) has a diameter of \SI{55}{\micro\meter} and the satellite droplet in C) has a diameter of \SI{21}{\micro\meter}. B) \& D) False color images probability of target presence for primary and satellite droplet targets. Note that the large gradient, as compared to Figure \ref{fig:columntarget}B), indicates a decrease in the dimensional and positional stability.}
	\label{fig:droplettargets}
\end{figure*}

Reduced mass targets have been explored for their uses in the study of LPI and warm dense matter due to enhanced electron refluxing and heating, resulting from the limitation of return currents which occur in bulk targets \cite{Bell1997}. Solid, metal-based reduced mass targets are relatively expensive and difficult to employ in LPI studies when compared to non-reduced mass targets. This is due to the added constraints imposed during fabrication of limited transverse dimensions and support of the target by narrow mounting wires. 

Liquid droplets are a promising alternative to the solid-based reduced mass targets, conventionally used. Prior work has been conducted on ion acceleration and subsequent neutron generation from heavy water droplets \cite{schnurer2004explosion} synchronized to the laser \cite{Karsch2003}. In related fields, metal droplet targets are commonly used for EUV and XUV generation, abiet at laser intensities far below the relativistic limit. In this section we present a reduced mass target based on liquid droplets which are ideal for high repetition rate studies with highly repeatable size and positional control.

As previously addressed, the inherent instability of the liquid jet causes a breakup into droplets after a given distance. This disintegration of the liquid jet is driven by minimization of the surface energy of the fluid and initiated primarily by vibrations and sheer stresses within the liquid jet. Here we intentionally seeded  the instability of the jet, via the Plateau-Rayleigh instability, to create droplet formation with high repeatability. This causes the formation of droplets at frequencies greater than \SI{100}{\kilo\hertz} with high precision volume and velocity distributions. 

Seeding of the Plateau-Rayleigh instability requires a vibrational or pressure perturbation to be applied to the liquid jet. For this work we seed a vibrational instability with a Thorlabs AE0203D08F piezoelectric actuator which is affixed to the nozzle nut with epoxy as shown in Figure \ref{fig:nozzle_and_jet}. When operated near the spontaneous droplet formation frequency, due to the resonance properties of this effect, a small amplitude, few cycle vibration generated by the actuator is sufficient to reinforce instability growth in the liquid microjet.

In high repetition rate use, synchronization between the fixed laser pulse frequency, \SI{0.0167}{\hertz} - \SI{1}{\kilo\hertz}, is required in order to have positional stability of the droplet with respect to the laser focus. We employ a \SI{1}{\kilo\hertz} trigger signal synchronized to the laser source to time the actuator driver with variable drive frequency, pulse number, pulse duration, delay, and amplitude. While the trigger signal arrives every \SI{1}{\milli\second}, the actuator driving signal is run in a burst mode with a frequency near that of the spontaneous droplet formation frequency ($f_d =$ \SI{178}{\kilo\hertz}). Illumination by the shadowgraphy probe pulse, synchronized to the 1 kHz illumination laser source, verifies the timing and stability of droplet formation (Figure \ref{fig:droplettargets}).

In the process of droplet break off, large primary and small satellite droplets are formed in an alternating droplet train which is depicted in Figure \ref{fig:hydrodynamics}. For the \SI{30}{\micro\meter} diameter capillary, the large droplet, pictured in Figure \ref{fig:droplettargets}A) has a diameter of \SI{55}{\micro\meter}. Formation of the satellite droplets within the droplet train does not occur for all conditions and is dependent on the viscosity and surface tension of the fluid \cite{Pimbley1977}. For water, used here, satellite droplets shown in Figure \ref{fig:droplettargets}C) are formed which alternate with the primary droplets along the propagating droplet train. The satellite droplets are notably smaller than the orifice diameter at just \SI{21}{\micro\meter} in diameter.

The shape of the primary droplets are slightly ellipsoidal. The major axis length is \SI{56}{\micro\meter} and the minor axis length is \SI{54}{\micro\meter}. The $1\sigma$ standard deviation of the size of these axes is less than $\pm$\SI{1}{\micro\meter} from droplet-to-droplet. 

While the dimensional measurements are precisely controlled, the positional stability of the primary droplet target is relatively less well constrained as illustrated by the probability of target presence map shown in Figure \ref{fig:droplettargets}B). For this case, the standard deviation of the centroid position in the horizontal plane, $\sigma_x$, is \SI{1.5}{\micro\meter}. In the vertical plane, the centroid standard deviation, $\sigma_y$, is \SI{5.7}{\micro\meter}. 

The smaller satellite droplet is also slightly ellipsoidal in shape with a major axis length of \SI{22}{\micro\meter} and minor axis length of \SI{20}{\micro\meter}. The dimensional $1\sigma$ values for both axes are less than \SI{1}{\micro\meter}. The centroid positional stability is more well constrained for this droplet type. Here $\sigma_x$ is less than \SI{1}{\micro\meter} and $\sigma_y$ is \SI{2.4}{\micro\meter}.

The dimensional and positional stability of the droplet targets is critical for use, especially in the case of reduced mass targets. The droplet targets demonstrated here are well-suited for use with fast focusing optics as the positional stability is better than the length of the confocal parameter even for an $f/1$ focusing optic. It may be of interest that, sub-five micron droplet generation is possible using smaller diameter capillaries and associated additional complications. Further improvements in the droplet positional stability may be made in future version of the target system which are designed to optimize this parameter. 

A number of other approaches to generating droplets have been performed. These methods include pressure-based initiation of the Plateau-Rayleigh instability as opposed to vibrational \cite{Martin2008}. Cylindrical piezoelectric actuators which surround the capillary have been shown to be a repeatable method of triggering droplet formation \cite{Perduijn1983}. Even lasers have been used to perturb a liquid microjet and generate repeatable droplet formation \cite{chvykov2010microdroplet}.

\subsection{Liquid Sheet Targets}
Planar, solid density, foils of order a few microns thick are the most commonly employed target configuration for the study of high intensity LPI. These foils have proved useful for the study of a wide range of processes including: energetic electron and ion acceleration \cite{TNSA_Hatchett}, x-ray generation \cite{Yu1999}, and even ultra-intense high harmonics \cite{Dromey2006}. For high repetition rate studies and applications, a liquid target with planar geometry and submicron thickness is required.

Literature on the formation of liquid sheets is abound, stemming from range of fields \cite{hasson1964thickness,choo2001parametric,bush2004collision}. The requirements which we have outlined, however, have yet to be satisfied. As a result, we build on these other works in order to meet the needs required for use in LPI.

More recently, contemporary efforts to create a flowing, planar, liquid sheet target have resulted in the development of two approaches. First, the method upon which this work is based, is the intersection of two, laminar flowing liquid microjets. Previous work by Ekimova \textit{et al.} 2015, demonstrated the formation of liquid sheet targets as thin as \SI{1.4}{\micro\meter} in vacuum through the head-on intersection of two \SI{50}{\micro\meter} diameter water jets \cite{Ekimova2015}. Our work improves upon this result in two ways: use of non-normal incidence between the two microjets results in the generation of sheets as thin as \SI{450}{\nano\meter} and operation with ethylene glycol significantly improves the vacuum compatibility of the system. 

The second method is through the use of microengineered nozzles. Galinis \textit{et al.} 2017 use 3D printed nozzles with \SI{200}{\nano\meter} resolution to construct a tapered nozzle orifice which is \SI{260} by \SI{30}{\micro\meter}. This forms a sheet as thin as \SI{1.49}{\micro\meter} which is shown to be stable in vacuum and at atmospheric pressure \cite{Galinis2017}. Another method, by Koralek \textit{et al.} 2018, uses microfluidic gas-dynamic nozzles to produce liquid sheets from \SI{1}{\micro\meter} to \SI{10}{\nano\meter} \cite{Koralek2018}. This is achieved through pinching of a central microjet by two impinging gas jets.

\begin{figure}[h]
	\centering
	\includegraphics[width=7.7cm]{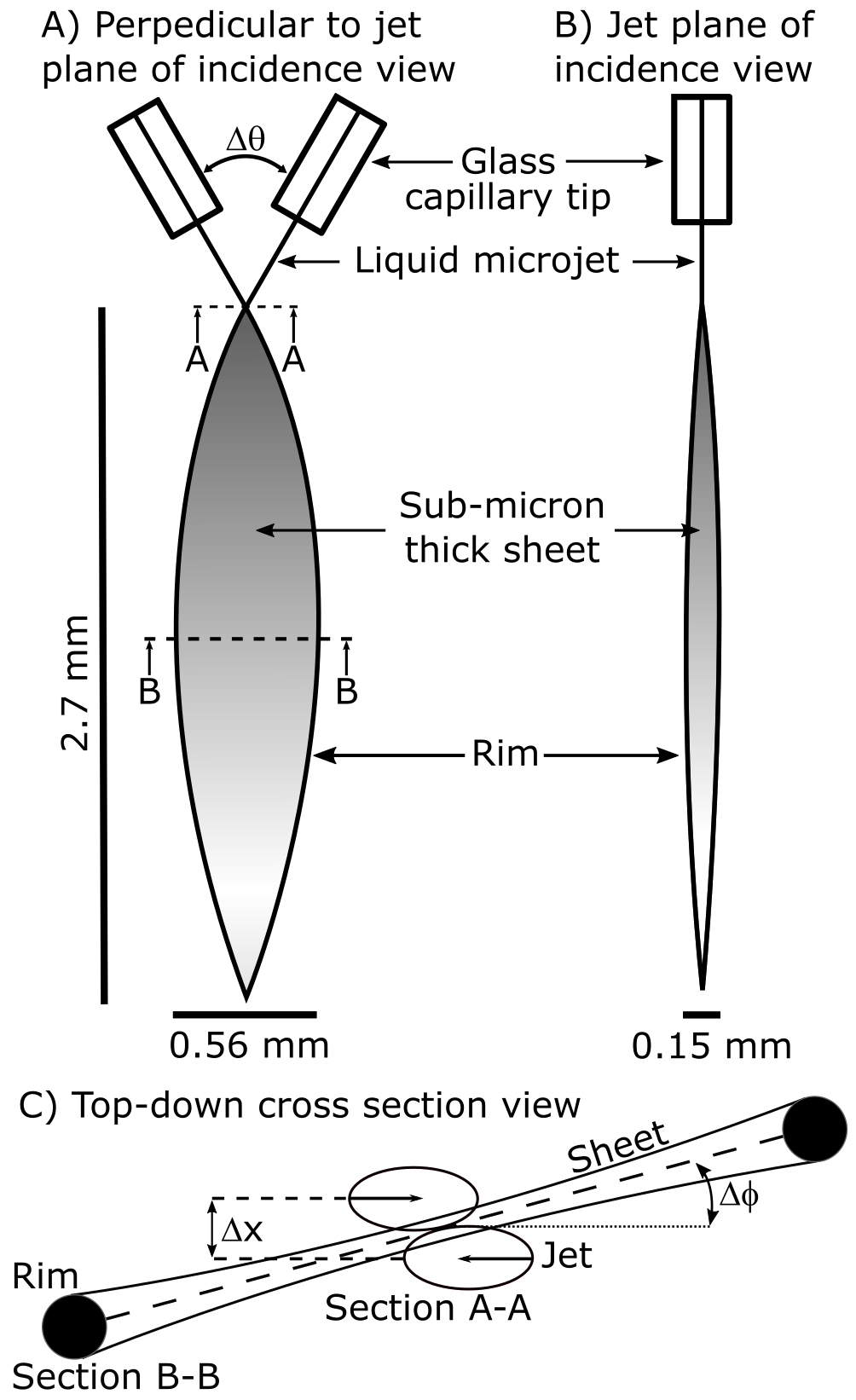}
	\caption{A) View of capillary nozzles and thin liquid sheet formed perpendicular to the plane of incidence between the jets. The full angle between the two jets is denoted $\Delta \theta$. B) View of the sheet formation within the jet plane of incidence. Note that the sheet is not aligned to this plane due to the grazing incidence of the two jets. C) Top-down cross section views of the jet intersection and resulting sheet formation. Section A-A illustrates the grazing incidence of the two jets which allows the formation of a submicron thick sheet. Section B-B shows the relative angle, $\Delta \phi$ of the plane of the sheet with respect to the plane of incidence of the jets. The thick, cylindrical rim which supports the sheet is shown as well.}
	\label{fig:sheet_geometry}
\end{figure}

The nozzle arrangement for this work and the resulting target geometry is illustrated in Figure \ref{fig:sheet_geometry}. Through the introduction of a second nozzle assembly, we generate planar, submicron thick, liquid sheet targets. In the impingement of two equal diameter, equal velocity liquid microjets, a leaf shaped, thin sheet is formed. With the use of \SI{30}{\micro\meter} diameter capillaries and ethylene glycol, the sheet is less than \SI{1}{\micro\meter} thick and displays high dimensional and positional stability. The high-velocity laminar flow additionally makes it suitable for high repetition rate use at greater than \SI{10}{\kilo\hertz}, as the ablated interaction region is refreshed every \SI{25}{\micro\second} for laser conditions used. 

To form the sheet, two \SI{30}{\micro\meter} diameter glass capillaries are aligned with the tips in close proximity. The full angle of incidence, $\Delta\theta$, between the two liquid jets is mechanically constrained to be \SI{60}{\degree}. Unlike other approaches, the jets are intentionally intersected with a grazing incidence where the degree of overlap of the two jets is precisely controlled by means of piezoelectric translation of one capillary with respect to the other. The amount of overlap between the two jets, $\Delta \textrm{x}$, ultimately defines the minimum thickness of the sheet. At normal incidence the thickest sheet is formed, typically resulting in a minimum thickness of a few microns. While at grazing incidence configuration, $\Delta \textrm{x} > 0$, a thinner sheet is generated until it is no longer stable and the lower half of the leaf-like shape does not reconnect at the bottom. 

It is important to note that the angle of the sheet relative to the plane of incidence between the two microjets is dependent on $\Delta \textrm{x}$, the amount of overlap between the jets. This point is illustrated in Figure \ref{fig:sheet_geometry}C). When the jets are normally incident, the sheet is formed perpendicular to the plane of incidence. However, the thinnest sheet is formed through grazing incidence where the sheet is clocked to have a $\Delta \phi \approx$ \SI{15}{\degree} from the plane of incidence. When $\Delta \phi <$ \SI{15}{\degree} the resulting sheet is unstable, with the lower half of the leaf-like shape open at the bottom.

\begin{figure}[t]
	\includegraphics[width=\linewidth]{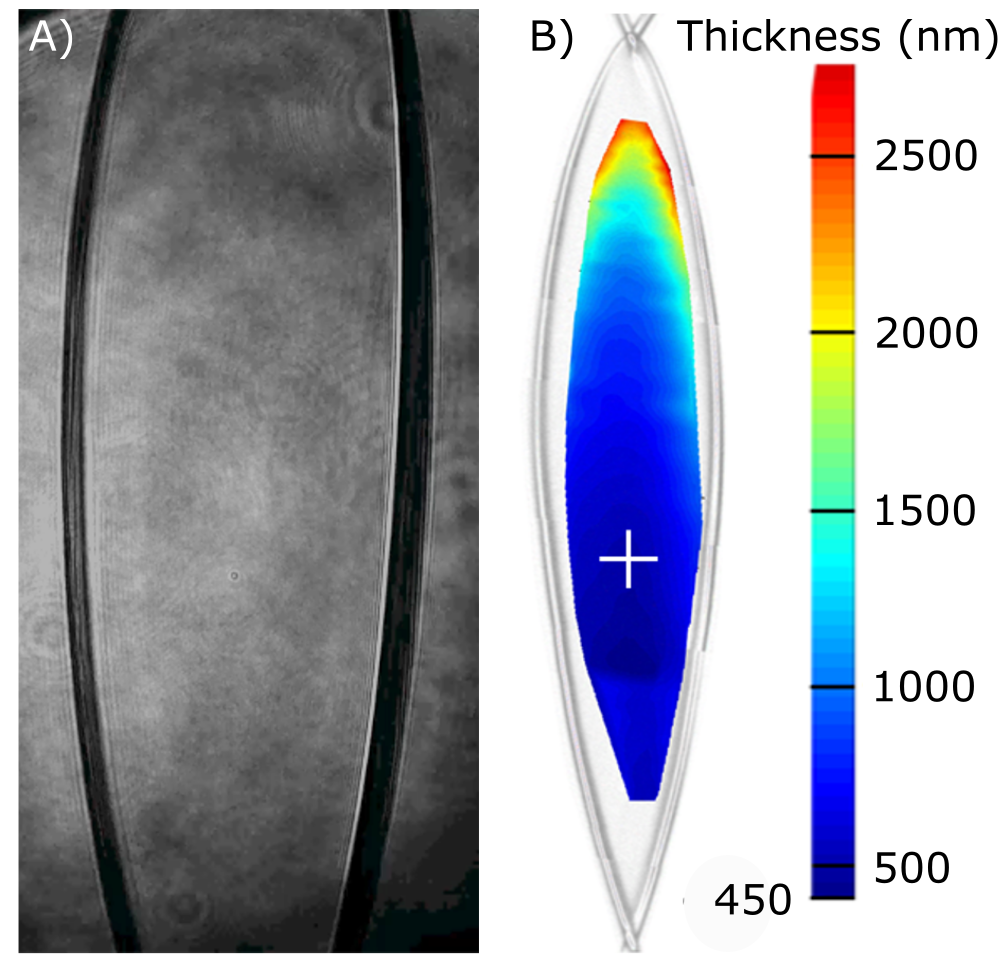}
	\caption{A) Microscope shadowgraphy image of the central region of the liquid sheet target in vacuum. B) Spatially dependent thickness map across the liquid sheet collected with Filmetrics white light interference profiler. The white cross indicates the location of the minimum sheet thickness at \SI{450}{\nano\meter}. This figure is reprinted with permissions from Morrison \textit{et al.} 2018 \cite{Morrison2018}.}
	\label{fig:thickness}
\end{figure}

Following the dual microjet geometry described above and with the use of ethylene glycol at a \SI{23.6}{\meter\per\second} fluid velocity and grazing jet impingement, the sheet formed is shown in Figure \ref{fig:thickness}A). The dimensions of the sheet are \SI{2.6}{\milli\meter} long by \SI{0.56}{\milli\meter} wide, as measured by microscope imaging. A two dimensional thickness map of the sheet was measured using a commercial Filmetrics white light thin film interference device and is shown in Figure \ref{fig:thickness}B). 

The sheet thickness at the top is a few microns thick. Further down, the sheet thins to a minimum of \SI{450}{\nano\meter} as denoted by the white cross in Figure \ref{fig:thickness}B). Progressing towards where the rims reconnect at the bottom of the sheet, the sheet thickness increases again.

For the above described configuration, with the use of ethylene glycol, \SI{450}{\nano\meter} was the minimum achievable thickness. It should be noted, however, that we have created sheets with water as thin as \SI{275}{\nano\meter} using the same jet configuration. Unfortunately, these sheets are unstable and the rims do not close at the bottom. Efforts to further reduce the sheet thickness are ongoing.

The structure of the sheet is supported by the two thick, $>$\SI{25}{\micro\meter} diameter jets which do not coalesce into the thin sheet and support the leaf-like shape. As a result, the edges of the sheet are relatively thick compared to the thin film at the center. 

Subsequent secondary and higher-order leaf-like structures are formed below the primary sheet. These sheets, however, are relatively thick in comparison to the first sheet and smaller in length and width. The sheet ultimately disintegrates into a droplet spray after the onset of the Plateau-Rayleigh instability.

Other notable variables of the submicron sheet target include the control of fluid velocity. As the fluid velocity increases, so does the length and width of the sheet. There are, however, practical limitations to the overall size of the sheet as determined by the PSI rating of the syringe pump, Reynolds number limit for laminar flow, and resulting spontaneous breakup length which is flow velocity dependent. Additionally, all of these values ultimately depend on the fluid used.

The sheet target is not depicted in terms of the probability of target presence, as it is not well characterized in terms of dimensional and positional stability by this illustration. The critical stability values for the sheet target are instead the target angle, optical axis positioning, and sheet thickness. The target angle stability was measured by means of image analysis from frames collected in the testing of the plasma mirror (see Section \ref{liquidoptics}). Thresholding and centroid identification was performed for the frames collected by the CCD camera from the specular, high-field reflection off the liquid sheet as a plasma mirror. The $1\sigma$ standard deviation in the reflected beam position centroid was less than 1 pixel in both $x$ and $y$. The longitudinal positional stability of the sheet was measured to be better than \SI{2}{\micro\meter} by means of side-on microscope imaging. Lastly, the thickness stability of the sheet was found to be stable to better than \SI{3}{\nano\meter} over a \SI{10}{\micro\meter} patch as evidenced by the etalon-like thin film interference measurement performed in Section \ref{liquidoptics}.

The above described submicron thick, planar liquid sheet target has already been demonstrated for use in high intensity, high repetition rate LPI experiments. Morrison \textit{et al.} 2018 employed this target, in combination with a kHz repetition rate, millijoule-class, relativistically intense laser to generate energetic protons at up to \SI{2}{\mega\electronvolt} at kHz repetition rate \cite{Morrison2018}. Previous efforts to accelerate ions at kHz repetition have relied on front surface TNSA from relatively thick targets, providing diminished efficiencies compared to rear surface TNSA \cite{Hou2009}. This demonstration is a substantial advance towards utilizing the full capability of high repetition rate lasers and meeting the needs of promising applications \cite{palmer2018paving}.

\subsection{Exotic Liquid Targets}

\begin{figure}[t]
	\includegraphics[width=\linewidth]{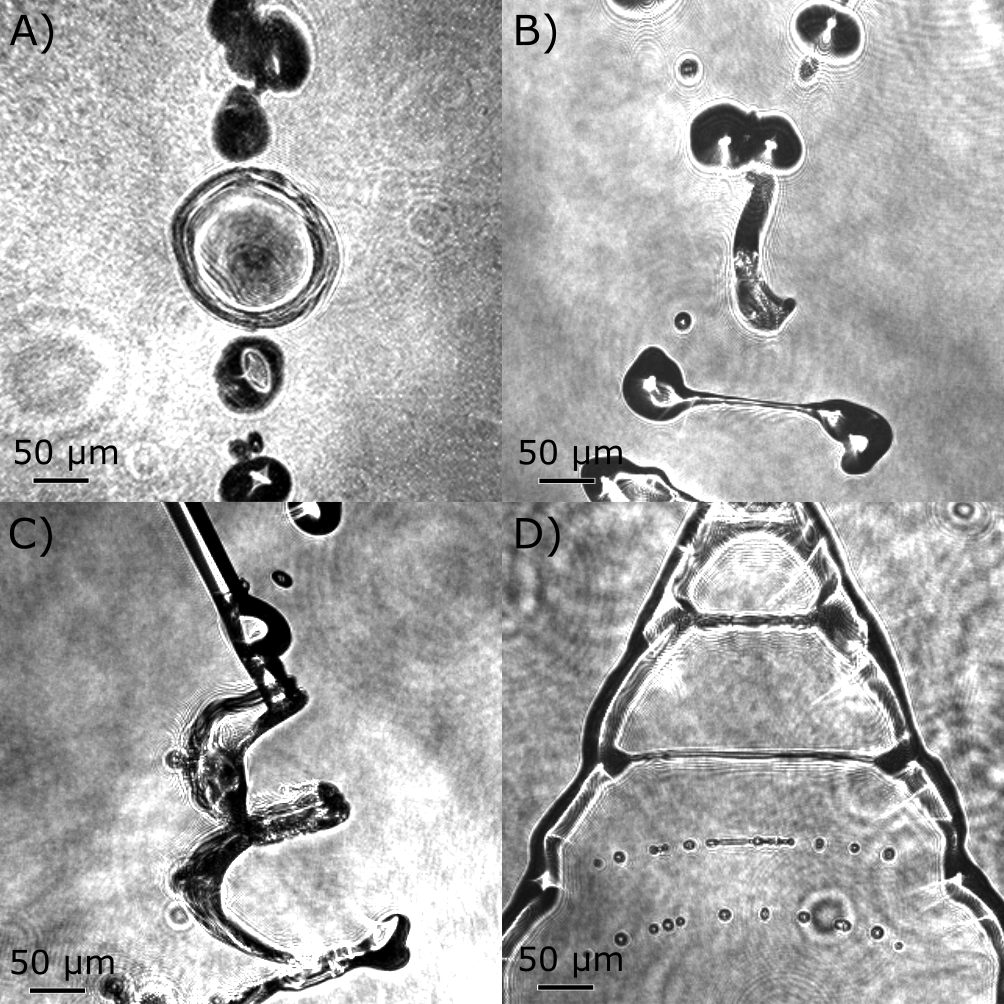}
	\caption{A variety of other unique target configurations can be created with droplets and jets. A) Face on view of droplet-droplet collision designed to make an isolated disk target. B) Side view of droplet-droplet isolated disk target shown in A). C) Droplet-jet collision generating a target with cylindrical surface shape. D) Thin ($\approx \SI{5}{\micro\meter}$ diameter) horizontal wire formed through the intersection of two jets while driving the Plateau-Rayleigh instability with piezoelectric device.}
	\label{fig:unique_targets}
\end{figure}

Aside from the relatively simple cylindrical, spherical, and planar geometries formed by the jet, droplets, and sheet targets, more exotic and complex geometries are possible. During our efforts we explored a few exotic configurations which may be of interest to the LPI community. These represent only a small subset of possible liquid targets. The results we highlight here are meant to be illustrative, not exhaustive. In this section we present isolated disks, cylindrically curved sheets, and few-micron diameter narrow wires as shown in Figure~\ref{fig:unique_targets}.

These targets have a range of potential use cases for both fundamental studies and applications. In particular, the isolated disks function as reduced mass targets. As previously addressed, reduced mass, planar targets are of high interest to the high intensity laser-plasma community for their known role in the enhancement of ion acceleration due to the promotion of enhanced electron refluxing and sheath fields \cite{Nilson2009,Mackinnon2002}. This results in higher conversion efficiency and peak ion energies for the TNSA ions. The cylindrically curved surface targets enable the potential for control of ion beam divergence. Previous work has demonstrated the use of curved surfaces in LPI experiments as a method to focus ion beams for secondary target heating \cite{Offermann2010}.

\subsubsection{Isolated Disk Targets} \label{isolatedpancake}

An alternative approach to producing thin planar targets, from the method detailed for the submicron thick sheet targets, is through the collision of two droplets \cite{Pan2009}. To create isolated disks, our two liquid jet nozzles are operated in droplet mode by oscillating the piezoelectric actuator near the spontaneous droplet frequency with a burst frequency of \SI{1}{\kilo\hertz} such that the droplet train is synchronized to the imaging probe pulse. One nozzle position was fixed while the other was translated in order to overlap two droplets just after break off from the liquid jet. The full angle between the two colliding droplets was \SI{60}{\degree}. The face-on and side views of the isolated disk target is shown in Figure \ref{fig:unique_targets} A) and B). A \SI{130}{\micro\meter} diameter disk is formed with a relatively thick rim with average diameter of \SI{17}{\micro\meter}. Using volume conservation from the two droplets, along with the thickness of the rim, the thin sheet spanning the center of the isolated disk is approximately \SI{8}{\micro\meter} thick which is certainly within the desired range for TNSA.

Tuning of the angle normal to the disk surface is performed through off normal collisions of the two droplets in the horizontal and vertical planes. The off-normal intersection serves to rotate and oblate the disk, though global rotation of the two nozzles can maintain the sheet symmetry while also achieving a rotation. Another tuning parameter which can be used to modify the interaction is the evolution time of the droplet collision. This changes the aspect ratio, shape, and general morphology of the droplet collision and could be exploited to create other unique target types.

Additionally, relative fluid velocity, angle of incidence, and fluid properties such as surface tension and viscosity modify the droplet interaction and the resulting disk size, inner sheet thickness, and subsequent droplet temporal evolution. These parameters have not been surveyed in this work, but are proposed for future studies with relevance to high intensity LPI targetry. 

Droplet-on-demand generators where the droplets are expelled from the capillary orifice only when requested, instead of the continuous droplet generation approach used here, may be applied for droplet-droplet collisions on lower repetition rate laser systems or those which have concerns about vacuum pumping rate or excess background gas pressure \cite{Switzer1991}.

\subsubsection{Cylindrically Curved Sheet Targets} \label{curvedsurface}

One method of generated a high repetition rate liquid target with a curved surface is through a droplet-jet collision. As shown in Figure \ref{fig:unique_targets} C), the droplet-jet collision forms a cylindrically shaped sheet with the primary axis of curvature oriented horizontally. At early times in the interaction, a saddle-shaped feature is formed with curvature along the vertical axis due to the diameter of the droplet (\SI{55}{\micro\meter}) being larger than that of the jet (\SI{30}{\micro\meter}). The overfill of the interaction streams past the jet until surface tension pulls back the fluid.

Further modification and tailoring of the curved surface target can be performed by varying the relative size of the jet and droplet, angle of incidence, fluid properties, etc. These variations should enable variation in the radius of curvature, thickness, and other relevant parameters of the resulting curved surface.

\subsubsection{Narrow Wire Targets} \label{narrowwire}

Here, the piezoelectric actuators are operated continuously near the spontaneous droplet frequency to establish the Plateau-Rayleigh instability. Before droplet break off, a modulation in the diameter of the jet is formed resulting from the resonant instability. When two peaks from this modulation are overlapped between the two jets, the collision forms a triangular, ladder-like structure. The horizontally oriented rungs shown in Figure \ref{fig:unique_targets} D) are as small as \SI{5}{\micro\meter} in diameter. The length of each rung is over \SI{200}{\micro\meter}, generating a long aspect ratio, narrow wire spanning two relatively thick jets.

Proposed uses for these exotic target types are not directly clear, but unique and novel geometries are commonly used in LPI studies to measure, enhance, or modify various parameters \cite{Morrison2012,Kar2016}. The exotic targets shown here are just a few of the numerous target configurations capable of being made with the liquid microjet assembly. The overall parameter space for liquid targets is far too broad to be addressed in detail in this work, but we hope that these unique configurations stimulate the community to consider the possibilities this technique presents.

\section{Liquid Optics} \label{liquidoptics}

Liquid-based optical elements are commonly used in a wide range of optical applications. Dye jet lasers, liquid lenses, and an array of various liquid crystal-based optics, including phase and amplitude modulators, prisms, and lenses are now ubiquitous. In many applications, use of fluids instead of more conventional solid-state optics offer performance benefits such as variable focal lengths, electrically addressable control, or consumable modes of operation.

Liquid optics for high repetition, high intensity LPI also show promise to offer advantages over conventional optics, primarily in cases where the optical element is consumable. Liquids offer the capability for rapid refreshment and low cost per shot, such that use at high repetition rates viable. By generating the optic on an individual shot-to-shot basis, this avoids the usual concerns that the optics will be damaged by the fluence of the laser pulse. This quality is particularly advantageous when employed in extreme environments such as those in the vicinity of the LPI.

\subsection{Liquid Plasma Mirror}

With the push to develop high intensity lasers which operate at kHz repetition rates or higher, associated optical devices must also meet these demands. One such class of optical devices aims to improve the temporal pulse contrast of the laser pulse by suppressing or removing prepulses and pedestal features that prematurely damage the target and generate preplasma which can be detrimental to experimental objectives such as high energy ion acceleration. Many solid state temporal pulse cleaning devices such as Pockels cells, saturable absorbers, crossed polarized wave generation (XPW), and optical parametric amplifiers (OPA) have been demonstrated at high repetition rates as well. One commonly employed temporal pulse cleaning technique, plasma mirrors, however, are not ideally suited for high repetition rate use due to the consumable nature of the mirror media.

Typically composed of an anti-reflection coating on an optical quality substrate, a plasma mirror maintains low reflectivity until the leading edge of the pulse generates a highly reflective plasma on the surface. This technique nominally results in a contrast enhancement by a factor of 100 at the expense of approximately $25\%$ of the energy. Since the irradiated region of the optic is destroyed on each shot, the substrate is rastered, realigned, and ultimately discarded after a series of exposures.

Previous demonstrations of liquid based plasma mirrors have been performed with the use of ethylene glycol flowing from a dye laser jet \cite{Backus1993}. While capable of operating at kHz repetition rate, the system exhibited poor contrast enhancement unless operated at Brewster's angle where a 400:1 contrast improvement was shown, at the expense of low reflectivity of only $38\%$ due to oblique incidence with \textit{P} polarization. Another liquid based plasma mirror technique is based on laminar fluid flow over circular apertures \cite{Panasenko2010}. This work showed a high reflectivity of $70\%$, but low contrast enhancement of $35$ due to the lack of anti-reflection properties. One of the most promising approaches utilizes variable thickness, submicron thick, liquid crystal films to create plasma mirrors \cite{Schumacher2017,poole2016plasmamirror}. These films exhibit anti-reflection properties due to thin film destructive interference when formed to the appropriate thickness, but they have only been demonstrated at repetition rates up to a few Hz. 

As described in the remainder of this section, we expand on these techniques to demonstrate the use of a less than \SI{1}{\micro\meter} thick liquid sheet formed by the intersection of two \SI{30}{\micro\meter} diameter, laminar-flowing liquid microjets of ethylene glycol to create a plasma mirror. The liquid sheet exhibits variable low-field reflectivity between $18.5$ and as low as $0.1\%$ due to etalon-like constructive and destructive interference. As a plasma, the reflectivity is $69\%$, providing a temporal pulse contrast enhancement of $690$. The presented liquid plasma mirror is demonstrated at a \SI{1}{\kilo\hertz} repetition rate in vacuum with an ambient pressure of less than $1$ millitorr.

\subsection{Anti-reflection properties of thin liquid sheet}

\begin{figure}[t]
	\includegraphics[width=\linewidth]{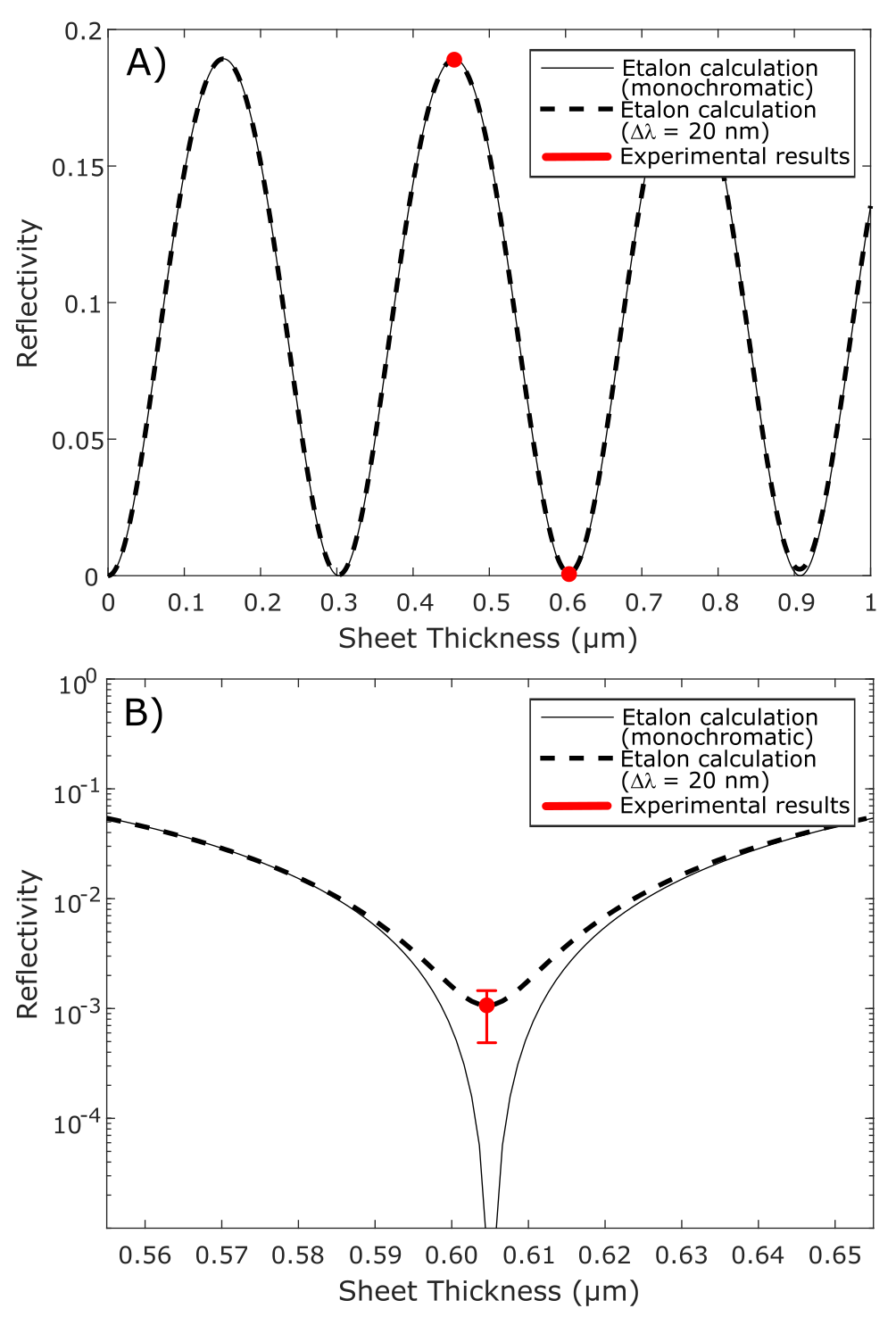}
	\caption{A) Etalon reflectivity as a function of thickness for the given experimental conditions. The single wavelength calculation is plotted with a solid line while the wavelength broadened curve corresponding to a Gaussian FWHM bandwidth of \SI{20}{\nano\meter} is given by the dashed curve. B) The bandwidth dependent etalon reflectivity is plotted in semi-log for the third minima to illustrate the effect of incidence with a broad bandwidth laser pulse. Note that while the etalon calculation continues towards zero at the minima for the monochromatic case, the minimum reflectivity for a pulse with \SI{20}{\nano\meter} of bandwidth is approximately $0.1\%$.}
	\label{fig:etalon_figure}
\end{figure}

Enhancing contrast with the use of a plasma mirror requires that the optical surface in the low-field exhibit low reflectivity. While solid-state plasma mirrors rely on dieletric coatings to achieve this effect, an alternative solution is to leverage an etalon-like effect of destructive interference in a thin film. The etalon reflectivity, $R_{\textrm{etalon}}$, for a thin film is given by :
\begin{equation}
R_{\textrm{etalon}} = 1 - \frac{1}{1+F\sin^2(\Delta/2)}
\label{eq:etalon_reflectivity}
\end{equation}
where
\begin{equation}
\Delta = \frac{4\pi n d \cos(\theta_\textrm{t})}{\lambda}
\end{equation}
and
\begin{equation}
F = \frac{4R_\textrm{i}}{(1-R_\textrm{i})^2}
\end{equation}
for the index of refraction $n$, sheet thickness $d$, internal angle of transmission in the film $\theta_{\textrm{t}}$, wavelength $\lambda$, and the polarization dependent single interface reflectivity $R_\textrm{i}$. For this work, we use ethylene glycol where $n = 1.4263$ \cite{Sani2014}, an external angle of incidence of \SI{35}{\degree} which gives $\theta_{t} = 23.71^\circ$, central wavelength $\lambda =$  \SI{790}{\nano\meter}, and $R_\textrm{i}$ defined for \textit{S} polarized light. The resulting etalon reflectivity as a function of film thickness is as shown in Figure \ref{fig:etalon_figure}A).

Note that the etalon reflectivity is a function of incidence angle and wavelength. By focusing onto the thin film, a range of angles are introduced. To mitigate this effect, we operate near focus, within the Rayleigh range, where the wavefronts are nominally flat. Therefore, we do not consider the angle dependent reflectivity and instead focus on the unavoidable wavelength dependence required to support an ultrashort pulse.

A MATLAB script was written to calculate the thickness dependent etalon reflectivity over the range of wavelengths contained within the pulse spectrum. This reflectivity was then weighted to the energy contained within each wavelength and is referred to as the bandwidth dependent calculation. To match the experimental laser pulse, a spectrum with Gaussian distribution centered at \SI{790}{\nano\meter} and full width half maximum $\Delta \lambda =$ \SI{20}{\nano\meter} was used for the bandwidth dependent calculation and is co-plotted with the reflectivity for the monochromatic case in Figure \ref{fig:etalon_figure} A) and B). 

The linear scale plot in Figure \ref{fig:etalon_figure}A) does not reveal a large discrepancy between the monochromatic etalon calculation and bandwidth dependent etalon calculation. When plotted in semi-log about the third minima, given by Figure \ref{fig:etalon_figure} B), the differences are apparent. The monochromatic etalon calculation rapidly declines towards the minima with $R_{\textrm{etalon}} < 10^{-5}$, while the wavelength dependent etalon calculation reaches a lower limit of $R_{\textrm{etalon}} \approx 10^{-3}$.

The bandwidth dependent etalon reflectivity of the thin film puts a lower limit on the reflectivity at each minima. At higher order minima the minimum reflectivity increases. Therefore, operation at the lowest minima possible is desired, which in our case is at \SI{605}{\nano\meter}.

\subsection{Experimental Setup and Results}
\begin{figure}
	\includegraphics[width=\linewidth]{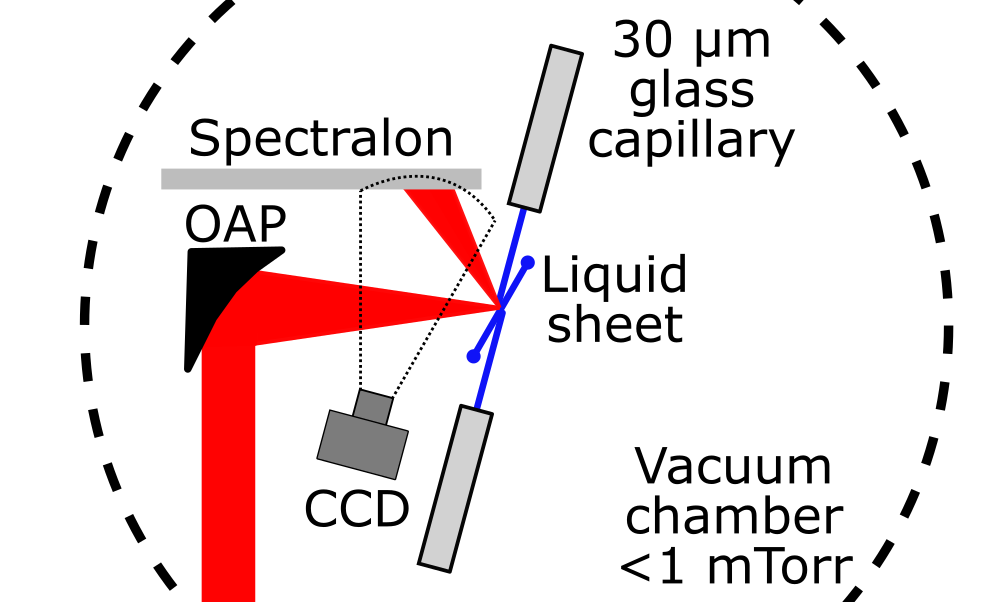}
	\caption{Experimental setup for measuring the thin, liquid sheet plasma mirror reflectivity. The laser was focused onto the liquid sheet at a \SI{35}{\degree} angle of incidence. The reflected light was then scattered by a Spectralon panel which was imaged by a CCD. While operating at 1 kHz repetition rate, the vacuum chamber pressure was maintained below 1 milliTorr.}
	\label{fig:experimental_setup}
\end{figure}

The thin, liquid sheet was experimentally tested for its anti-reflection properties and its use as a plasma mirror. A few hundred thousands shots were taken with \textit{S} polarized light at low intensity (I$<$\SI{e11}{\watt\per\centi\meter\squared}) to measure the reflectivity of the sheet in the low-field case. A schematic of the experimental setup is shown in Figure \ref{fig:experimental_setup}. A \SI{25.4}{\milli\meter} diameter, \SI{152.4}{\milli\meter} focal length protected gold coated off-axis parabolic mirror was used to focus pulses with a \SI{15}{\milli\meter} input diameter, \SI{50}{\micro\joule} of energy, and \SI{200}{\pico\second} pulse duration at \SI{1}{\kilo\hertz} repetition rate, onto the sheet at an angle of incidence of \SI{35}{\degree}. The focal spot size at the sheet was measured to be \SI{7.8}{\micro\meter} by \SI{12.1}{\micro\meter}, resulting in an intensity of approximately \SI{7.6e10}{\watt\per\centi\meter\squared}. Pump-probe shadowgraphy was used to confirm that the sheet was unperturbed by the laser pulse in the low-field. All tests for the plasma mirror were performed under vacuum at a chamber pressure below 1 millitorr.

\begin{figure}
	\includegraphics[width=\linewidth]{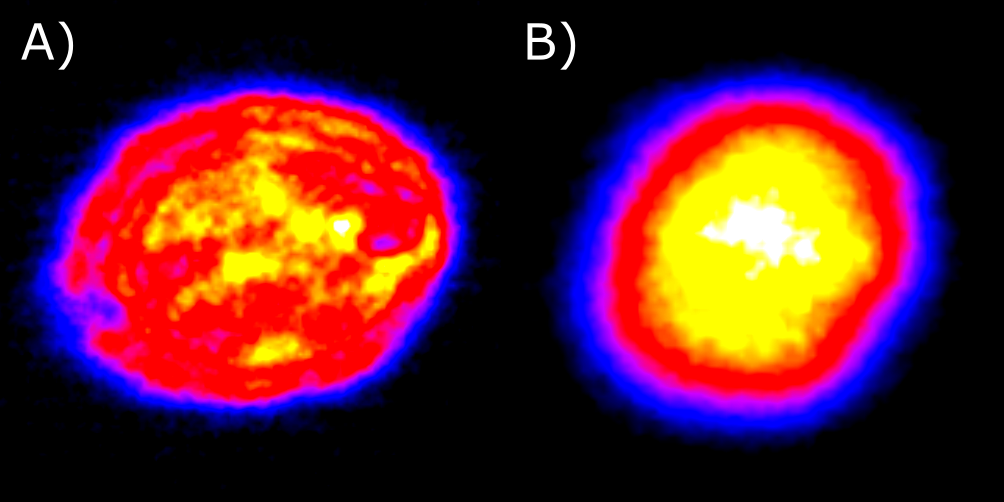}
	\caption{A) Near field mode of laser pulse input onto the plasma mirror. B) Near field mode of laser pulse after reflection from plasma mirror. Note the smoothing of the mode performed by the plasma mirror.}
	\label{fig:plasma_mirror_mode}
\end{figure}

In addition to temporal pulse contrast enhancement, plasma mirrors can exhibit spatial mode cleaning properties. When operated near the focus, as in this case, the higher order spatial modes are effectively filtered out of the reflected beam because the intensity dependent reflectivity is less in these regions. This effect was recorded in the high field with the input and reflected laser modes in the near field shown in Figure \ref{fig:plasma_mirror_mode}. While Figure \ref{fig:plasma_mirror_mode}A) exhibits many hard edges and diffractive features common in high intensity laser modes, the mode reflected from the plasma mirror, Figure \ref{fig:plasma_mirror_mode}B), smooths out these hard edge features resulting in a more smooth mode quality.

The reflectivity was recorded by means of an ImagingSource DMK23UP1300 CCD imaging a Spectralon panel. Calibrated filters were placed in front of the CCD and the camera integration time was changed to enable the high dynamic range measurement. Absolute calibration for the high and low-field measurements was performed by placing a protected silver mirror out of focus and reflecting the light directly onto the Spectralon panel.

Since the thickness of the sheet varies most drastically along its vertical length, from \SI{500}{\nano\meter} to greater than \SI{1}{\micro\meter}, the sheet was rastered vertically while observing the reflected light. In this way, the constructive interference peaks and destructive minima were surveyed and recorded. We observe a constructive interference maximum reflectivity in the low-field of $18.5\%$ and destructive interference minima of $0.1\% \pm 0.05\%$. These results are consistent with the constructive maxima shown in Figure \ref{fig:etalon_figure}. The destructive mimima was found to be congruous with the broad bandwidth etalon reflectivity calculation at the third reflectivity minima with a sheet thickness of \SI{605}{\nano\meter}. Though the sheet thickness was not measured in situ, the estimated value is well within the range of values provided by previous measurements performed with a Filmetrics commercial thin film interferometric measurement device.

Following the same procedure, the high-field reflectivity was recorded with the same experimental setup, but with \SI{1.3}{\milli\joule} of energy per pulse and a \SI{50}{\femto\second} Gaussian FWHM pulse duration, as measured by a single shot autocorrelator, for an on target intensity of \SI{8.8e15}{\watt\per\centi\meter\squared}. For these experimental conditions, a high-field reflectivity of $69\%$ was recorded. This value is in agreement with other results from literature for both solid-state and liquid-based plasma mirrors \cite{Ziener2003,Doumy2004,Thaury2007}. Incorporating the low-field reflectivity of $0.1\%$ with the high-field reflectivity of $69\%$, the contrast enhancement factor for this configuration is $690$.

\subsection{Discussion of Experimental Results}
Here, we have demonstrated, to our knowledge, the first liquid-based plasma mirror with etalon-like anti-reflection properties, capable of operating at repetition rates exceeding \SI{1}{\kilo\hertz}. The temporal contrast enhancement of 690 is comparable to, or exceeds results reported for both solid-state and liquid-based plasma mirrors operating at substantially lower repetition rates. This technique is highly stable and, with two high pressure syringe pumps, can be operated indefinitely.

These results, however, are not without drawbacks, which may limit the applicability of the described plasma mirror. The usable area on the plasma mirror, over which the low-field reflectivity is minimized and the sheet is locally flat, is approximately \SI{30}{\micro\meter} in diameter. Since the high-field reflectivity is a function of intensity, which peaks around $\approx$\SI{1e16}{\watt\per\centi\meter\squared}, the optimal incident energy for a \SI{30}{\femto\second} pulse is only \SI{2}{\milli\joule} \cite{Ziener2003,Cai2009}. Use with more energetic pulses over this region would result in lower reflectivity and similarly contrast enhancement. Alternatively, larger focal spot sizes would result in a reduced average low-field reflectivity and spatially dependent contrast enhancement, due to the varying thickness of the sheet over the focal spot size.

Operation of the plasma mirror at the third etalon minima also limits the bandwidth dependent low-field reflectivity. The ideal operational thickness condition is at the second minima, centered around \SI{300}{\nano\meter}. This lower minima permits a broader wavelength tolerance and lower low-field reflectivity over a larger range of sheet thicknesses.

Nonetheless, the results presented here establish a proof-of-principle demonstration of one such optical element commonly used in low repetition rate LPI studies, but now through the adoption of a fluid media, is adaptable for use in high repetition rate operation. Future developments in the creation of submicron thick liquid sheets will serve to further improve the range of laser parameters under which this type of device can be of use. Additionally, we hope to see the invention of other complementary liquid-based optical elements, such as the aforementioned plasma focusing optics.

\section{Practical Considerations}

As previously mentioned, we use ethylene glycol in the experiments presented in this paper. By commenting on desirable properties of liquids for these experiments and other constraints, this section provides some justification of this choice and other considerations for using other liquids for experiments of this kind. As will be discussed, for experiments requiring a vacuum environment, selecting a liquid with a low vapor pressure is important and there are other concerns for the design of the vacuum pump system to be considered, however, for experiments that do not require vacuum conditions a wider range of liquids can be employed (e.g. Morrison \textit{et al.} 2015).

As we anticipate the development of relativistically intense laser systems with repetition rates of kHz or higher, we estimate the maximum potential repetition rate for the cylindrical jet, droplet, and sheet targets. This measurement is performed by use of pump-probe shadowgraphy with high temporal resolution over microsecond time delays. With this technique, we determine the time required to reestablish a new target in the interaction region and infer the potential for use at repetition rates in excess of \SI{1}{\kilo\hertz}.

\subsection{Vacuum operation with liquid microjets} \label{vacuumoperation}
In the experimental study of high intensity LPI, vacuum operation is necessary in order to prevent nonlinear effects from degrading beam quality during propagation to the target \cite{Bliss1976,Mlejnek1998,Sun2004}, neutralization of ion acceleration \cite{Morrison2018}, and enable the use of electrostatic diagnostics such as a Thomson parabolic spectrometer \cite{Paschen1889}. Vacuum operation with the described liquid target system requires a two pronged approach. First and foremost is the use of the appropriate fluid with relatively low vapor pressure with the added complication of debris-free operation. Second is the efficient extraction or containment of excess liquid from the primary vacuum chamber where the laser-target interaction occurs. 

While the liquid target system is compatible with a number of fluids, those with low vapor pressure are ideal for low vacuum operation. Counter to this point is the added constraint of relatively low viscosity. In practical application, the fluid viscosity is limited only by the available pump pressure. Due to the choked flow nature of the long aspect ratio glass capillary orifices, the bias pressure required for flow is increased as compared to less restrictive nozzles. Use of scanning electron microscope apertures or tapered nozzles reduce this constraint, and enable the use of higher viscosity and lower vapor pressure fluids for the same bias pressure.

Another important fluid characteristic to consider is debris-free operation. The amount of debris generated with high repetition rate lasers, as compared to the low repetition rate systems currently used, is substantially increased. Metal-based targets, when used with low repetition systems, typically rely on the low number of laser shots or thin pellicles in order to avoid appreciable accumulation of ablated material from deteriorating or damaging sensitive optics and diagnostics. Liquid metals present low vapor pressures, but produce debris accumulation within the vacuum environment rendering them unsuitable for high intensity, high repetition rate work \cite{Fujioka2005,Takenoshita2004}. 

Use of the appropriate debris-free fluid, where excess target material is evacuated from the chamber as gas load on the vacuum pumps, significantly aids operation. Further, cleanup is expedited and removal of target material build up on delicate optics is rendered unnecessary.

\begin{figure} 
	\includegraphics[width=\linewidth]{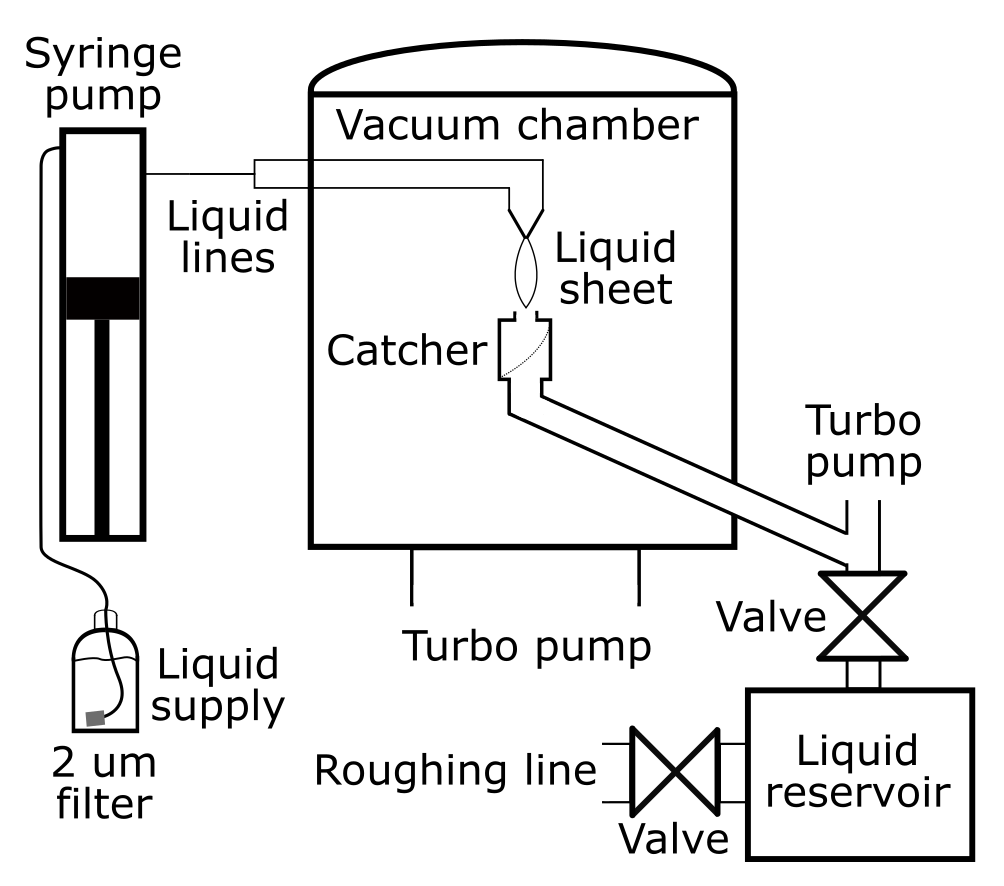}
	\caption{Schematic of vacuum and fluid containment system employed in this work. The syringe pump is fed by a liquid supply at atmospheric pressure. The supply line is capped with a \SI{2}{\micro\meter} sintered steel filter to prevent debris from entering the system. 1/16 inch Swagelok  lines and fittings are used to route the fluid into the vacuum chamber where the two nozzles generate the liquid sheet. A catcher and slide, designed to reduce gas backflow and splatter, direct the residual fluid to the liquid reservoir. Here the excess isolated from the main vacuum chamber by a turbomolecular pump. The reservoir is kept at relatively low vacuum pressure ($\approx 100$ millitorr) through backing with a roughing line. Liquid from the reservoir can be recovered and reused in the system.}
	\label{fig:fig10}
\end{figure}

In our work to improve vacuum levels during liquid sheet target operation, we learned fluid containment and extraction in the form of a proper fluid catcher design is crucial. Initial efforts to improve liquid collection and vacuum isolation of the main chamber from the fluid catch chamber used small aperture orifices. As the syringe pump is brought to pressure, droplets fall from the glass capillaries into the orifice, forming a meniscus. Once brought to pressure, the liquid microjet does not produce enough pressure to blow out the meniscus, and instead causes entrainment \cite{Bin1993,Qu2013}. The bubbles coalesce, expand, and sputter fluid back into the main vacuum chamber, disturbing the liquid sheet and raising the background chamber pressure.

Bubbling at the catch orifice was resolved by utilizing a relatively large, \SI{4}{\milli\meter} diameter orifice where the initial droplets fell through the aperture and did not form a meniscus. A slide was placed in the catcher below the orifice to intercept and guide the microjet stream further into the catch tube. Without the slide, uncontrolled splashing rapidly increases the available surface area for evaporation. This results in increased gas pressure within the catch tube generating a backflow of gas near the orifice, disturbing the sheet stability.

The complete vacuum system used in this work is illustrated in Figure \ref{fig:fig10}. Here, the liquid supply is maintained at atmospheric pressure before filling the syringe pump \SI{100}{\milli\liter} reservoir. Swagelok lines, with $1/16$ inch outer diameter, then route the fluid, under pressure, from the syringe pump to the nozzle assembly. Once the fluid is expelled from the capillary, it is recollected by the catcher orifice and slide, where it is redirected to a differentially pumped liquid reservoir. The pressure in the reservoir can be controlled through the valves to the turbo pump and roughing line. 

With the above described catcher orifice and slide implemented, a base pressure of $500$ microtorr was achieved with a \SI{2}{\milli\liter\per\minute} flow rate of ethylene glycol in the vacuum chamber. For reference, the turbo pump on the main vacuum chamber provides a pumping rate of \SI{2100}{\liter\per\second}. When the thin, liquid sheet is incident with a relativistically intense laser pulse at kHz repetition rate, the vacuum chamber pressure increases to $800$ microtorr. After this initial increase in pressure, the vacuum pressure is stable and operates continuously.

There is a threshold where the amount of ablated material from the laser-matter interaction cannot be effectively removed from the vacuum chamber. As a result, the chamber pressure increases above the ultimate base pressure, as in the case presented above. This resulting vacuum pressure may be too high to operate high voltage biased diagnostics such as Thomson parabolic spectrometers or to avoid nonlinear optical effects from impacting beam quality. Efforts to further improve the vacuum compatibility of the fluid or configuration of the liquid containment system are ineffective methods to decrease the ambient pressure under these circumstances since the ablated material from the LPI is the primary degrading cause.

Instead, one must deal directly with the amount of ablated material created. Some proposed methods to address this issue are: decrease the laser power through a reduction in pulse energy or repetition rate, increase the pumping rate, or implement reduced mass targets. The first approach works in direct competition to the desired high repetition rate and high average power operation of these laser systems. Increasing the pumping rate on the vacuum chamber is relatively straightforward, albeit somewhat costly. Use of reduced mass targets may present the most promising approach, in that it directly limits the potential amount of ablated material. These, and many other unforeseen issues of practical application must be addressed going forward with high repetition rate LPI studies.

\subsection{Repetition rate capability}

\begin{figure*}[t] 
	\includegraphics[width=\textwidth]{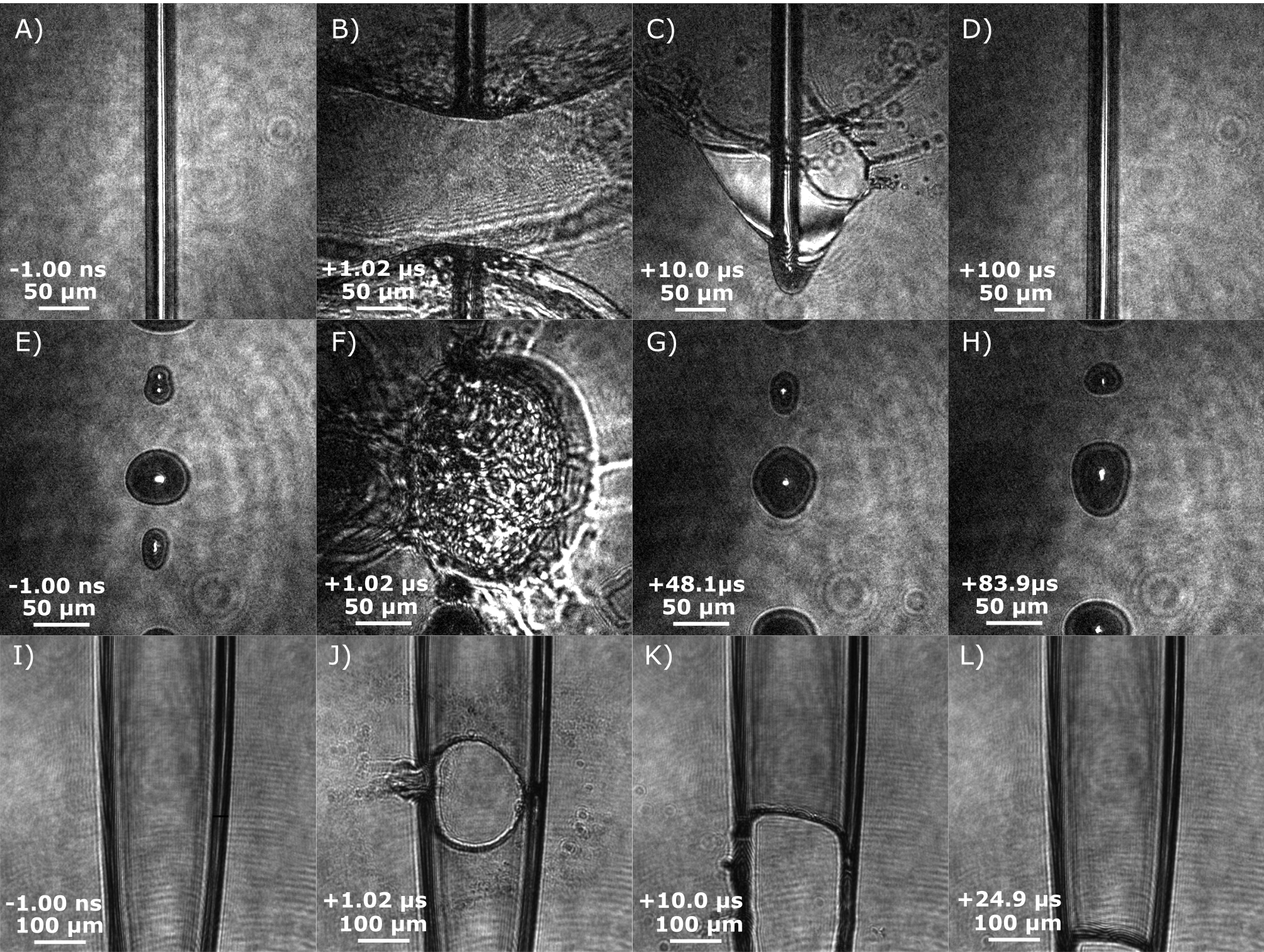}
	\caption{Short pulse (\SI{80}{\femto\second}) shadowgraphic microscope images of the hydrodynamic evolution before and after irradiation with a high intensity (\SI{e18}{\watt\per\centi\meter^2}) laser pulse for liquid column (A-D), droplet (E-H), and sheet (I-L) targets. These images illustrate the modes of target deformation and prescribe feasible repetition rates for each type. A) The liquid column begin as a continuous jet which is broken where the laser is incident, as shown in B). C) \SI{10}{\micro\second} after the laser arrives the column is reestablished, while a conical sheet and droplet spray obscure the laser line of sight. D) After \SI{100}{\micro\second} the droplet spray is cleared and a new column is set. E) A continuous train of primary and satellite droplets are shown before the laser arrives. F) Hydrodynamic explosion of the primary droplet destroys neighboring droplets within the series. In G) and H) separate primary droplets propagate into the appropriate position for subsequent laser shots after \SI{48.1} and \SI{83.9}{\micro\second}. I) Two colliding liquid jets form a thin, flowing sheet. J) \SI{1}{\micro\second} after the laser-target interaction a circular hole is vaporized. K) After \SI{10}{\micro\second} liquid flow begins to reform the sheet and at \SI{24.9}{\micro\second} the target surface is full reformed.   }
	\label{fig:hydrodynamics}
\end{figure*}

A practical consideration for application is the maximum possible repetition rate that these targets can support. In order to address this, pump-probe shadowgraphy was used to image the time evolution of the target geometry, along with the appearance of target debris in the region surrounding the laser-target interaction out to the \SI{100}{\micro\second} timescale. This evaluation is largely qualitative due to the lack of a relativistically intense laser with repetition rates exceeding \SI{1}{\kilo\hertz}. Nonetheless, we are unable to posit any deleterious effects which may occur at higher repetition rates aside from vacuum pressure deterioration due to additional material vaporization which is discussed in Section \ref{vacuumoperation}.

The following analysis is illustrated in the frames shown in Figure \ref{fig:hydrodynamics}. The laser parameters used here are an intensity of ~\SI{3e18}{\watt\per\centi\meter\squared}, pulse duration of \SI{40}{\femto\second}, and a pulse energy of \SI{5}{\milli\joule}. First, we assess the cylindrical column target formed by a single microjet shown in Figure \ref{fig:hydrodynamics} A-D). The initial liquid column is ablated by the laser pulse which vaporizes the region surrounding the laser-target interaction. This ablated target material forms a discontinuity in the microjet \SI{1}{\micro\second} after the interaction, which extends over \SI{100}{\micro\meter}. After \SI{10}{\micro\second} the microjet has propagated downward and an inverted umbrella-like shape forms which generates a spray of droplets obscuring a subsequent from cleanly interacting with the column. By $+$\SI{100}{\micro\second}, the fluid has propagated sufficiently downward to avoid this droplet spray and present a new target for the subsequent laser pulse to interact with. This indicates a potential for \SI{10}{\kilo\hertz} operation; a value within an order of magnitude of that found by Stan \textit{et al.} 2016 using of x-ray pulses \cite{Stan2016}.

The time evolution of the large primary droplet type is shown in Figure \ref{fig:hydrodynamics} E-H). 

Approximately \SI{1}{\micro\second} after the laser pulse deposits its energy, the droplet is still hydrodynamically expanding. Adequate time for the ablated material to evacuate the laser focus region is required.  This restricts the repetition rate capability of the droplets to only one out of every tens of droplets from the greater than 100 kHz droplet train. By +\SI{48}{\micro\second}, this debris has dissipated from the laser-target interaction region and a second droplet is ready to be shot. One can estimate the droplet targets are capable of operating at a repetition rate of up to \SI{20}{\kilo\hertz}. 

Lastly, we evaluated the sheet target as shown in Figure \ref{fig:hydrodynamics} I-L). In this geometry the umbrella-like spray from the column target, Figure \ref{fig:hydrodynamics} C), does not form. Instead, only the thin central region of the sheet must be reestablished. The continuity of the rim supports is illustrated by frame K). After \SI{25}{\micro\second}, the central, thin region of the sheet has propagated past the laser-target interaction spot resulting in a clear target for the following laser pulse. Thus, we conclude that the sheet target is suitable for nearly \SI{40}{\kilo\hertz} repetition rate for these laser conditions.

One practical point of discussion with regards to the repetition rate capabilities is that of scaling with pulse energy. As the pulse energy increases, the amount of material which is ablated and the resulting damage spot both increase. We expect that for Joule-class lasers the repetition rate capabilities of these target types will be lower than the few millijoule case which we present. In \SI{1}{\milli\second} the fluid will propagate \SI{2}{\centi\meter}, which is nearly an order of magnitude above the extent of the damage expected from Joule class lasers \cite{HecktandZajac1974}. While \SI{40}{\kilo\hertz} operation with a Joule-class laser may not be feasible, we certainly expect it to be appropriate for \SI{1}{\kilo\hertz} operation.

\section{Conclusion} \label{discussion}
We have described the use of fluids, based on high-velocity, laminar-flowing, liquid microjets, as targets and optics for the application and study of high intensity laser-plasma interactions at $\geq$ kHz repetition rate. Formed at room temperature by a robust and simple nozzle assembly, we configure the microjets to create cylindrical jets, droplets, submicron thick sheets, and several other unique targets. Short pulse shadowgraphy is used to characterize the targets and to illustrate their dimensional and positional stability.

Complementary to this effort, we demonstrate a consumable, liquid optical element in the form of a plasma mirror capable of kHz repetition rate operation. The mirror provides etalon-like thin film destructive interference with $0.1\%$ reflectivity for low optical intensities. At high intensities, where the mirror is in the plasma phase, this configuration produced $69\%$ reflectivity.

We discussed the practical implementation of either target or plasma mirror, including the compatibility of various microjet fluids with in-vacuum operation below 1 millitorr. Lastly, we illustrated through pump-probe shadowgraphy the repetition rate capabilities which exceed \SI{10}{\kilo\hertz}.

The above described targets and optics can be implemented in a wide range of future studies with scope beyond the field of high intensity laser-plasma interactions. The self-refreshing nature of the targets would present them as ideal for destructive or consumable operation in studies of high harmonic generation \cite{Dromey2009,Rodel2012,kahaly2013direct,quere2008phase,luu2018extreme}, shock dynamics \cite{fortov2007intense,tsuboi1994nanosecond,strycker2013femtosecond}, and X-ray free electron laser irradiation \cite{weierstall2014liquid,kim2010ultrafast,mafune2016microcrystal}. The submicron thickness of the sheet target configuration would suggest its use in research of soft X-ray spectroscopy \cite{smith2017soft,fondell2017time,kleine2018soft}, interfacial and surface chemistry \cite{ammann2018x}, and even positron scattering \cite{blanco2016scattering,tattersall2014positron}. Future research and development of liquid-based, vacuum compatible targets and optics will increase the already numerous potential applications.

The flowing-liquid targets and optics we presented are two pieces of a new, high repetition rate mode of operation for research involving high intensity laser-plasma interactions - a mode of operation meeting new demands from the development, construction, and availability of high repetition rate, relativistically intense lasers. Flowing-liquid targets and optics scale well to very high repetition rates, while providing densities higher than gas-based targets. Compared with using solids for the same purpose, flowing liquids present a largely debris-free, vacuum-compatible, and self-refreshing alternative. This enables relativistic LPI to generate quasi-continuous, high average flux sources of electrons, ions, X-rays, and neutrons for use in future application. For these reasons, we encourage the community to adopt this technique to move forward to new and exciting applications.

\section{Acknowledgements}
This research is supporoted by the Air Force Office of Scientific Research under LRIR Project 17RQCOR504 under the management of Dr. Riq Parra. Support was also provided by the AFOSR summer faculty program.

\bibliographystyle{unsrt}
\bibliography{bibliography}

\end{document}